\documentclass[letterpaper,aps,pra,preprint,showpacs]{revtex4}

\usepackage{graphicx}
\usepackage{braket}
\usepackage{amsmath}
\usepackage{amssymb}
\usepackage{dcolumn}
\usepackage{bm}

\newcommand{\lsp}{\ensuremath{\mathcal{L}_{\text{sp}}}}

\newcommand{\rhod}{\ensuremath{\dot{\rho}}}

\newcommand{\sigd}{\ensuremath{\dot{\sigma}}}

\begin{document}

\title{Predictions of laser-cooling temperatures for multilevel atoms in three-dimensional polarization-gradient
fields}

\author{Josh~W.~Dunn}
\author{Chris~H.~Greene}
\affiliation{JILA, University of Colorado and National Institute of Standards and Technology, and
Department of Physics, University of Colorado, Boulder, Colorado 80309-0440}

\date{\today}

\begin{abstract}
We analyze the dynamics of atom-laser interactions for atoms having multiple, closely spaced,
excited-state hyperfine manifolds. The system is treated fully quantum mechanically, including the
atom's center-of-mass degree of freedom, and motion is described in a polarization gradient field
created by a three-dimensional laser configuration. We develop the master equation describing this
system, and then specialize it to the low-intensity limit by adiabatically eliminating the excited
states. We show how this master equation can be simulated using the Monte Carlo wave function
technique, and we provide details on implementation of this procedure. Monte Carlo calculations of
steady state atomic momentum distributions for two fermionic alkaline earth isotopes, $^{25}$Mg and
$^{87}$Sr, interacting with a three-dimensional lin-$\perp$-lin laser configuration are presented,
providing estimates of experimentally achievable laser-cooling temperatures.
\end{abstract}


\maketitle

\section{Introduction}

The complex behavior that occurs when a multilevel atom interacts with polarization-gradient fields
has been of interest for some time now. Sub-Doppler cooling~\cite{phillips_88} occurs because of
elaborate optical-pumping processes produced by laser light in atoms with sublevel structure, as
seen, for example, in the lin-$\perp$-lin and the $\sigma_{+}$-$\sigma_{-}$ laser configurations.
The semiclassical understanding of these
interactions~\cite{cohen-tannoudji_89,chu_89,agresti_97,molmer_91,javanainen_92,javanainen_94} in
one or more dimensions has led to a reasonably good qualitative understanding of the underlying
mechanisms. Semiclassical analysis has even in some cases provided quantitative predictions of
sub-Doppler laser cooling temperatures measured in experiments~\cite{javanainen_94}.

However, the most direct route to a quantitative understanding of atom-laser interactions is via a
fully quantized master equation for the atom, in which the center-of-mass (CM) motion of the atom
is taken into account quantum mechanically. This allows behavior at low laser intensities and low
atomic velocities, the regime laser cooling strives to reach, to be described correctly. The
drawback of solving such a master equation, however, is the large number of basis states required
for the calculation, due to the additional momentum states. This problem becomes especially
pronounced when attempting to model three-dimensional (3D) systems, where the state space grows as
the cube of the number of one-dimensional momentum states needed.

The Monte Carlo wave-function (MCWF) technique, introduced in the early 1990's has allowed
significant progress to be made on the subject of atom-photon interactions in 3D as well as
lower-dimensional calculations. The MCWF technique is a simulation procedure for the master
equation that involves propagation of single stochastic wave functions, rather than density
operators, with random processes occurring at random intervals due to interactions with the photon
field that cause spontaneous emission. It has been shown that this method is equivalent to the
master equation in the limit of a large number of independent stochastic wave
functions~\cite{molmer_93}. The MCWF technique has been successfully utilized to calculate 3D
sub-Doppler laser cooling temperatures for atoms with Zeeman degeneracy in the ground and excited
states~\cite{molmer_95}.

The majority of the research done on laser cooling has involved essentially two-level systems,
consisting of a ground state and an excited state, which may or may not contain degenerate
sublevels. However, some investigations have explored atomic systems in which multiple distinct
excited states come into play. In particular, the use of bichromatic laser
fields~\cite{kazantzev_87,grimm_90} to cool three-level $\Lambda$ systems have been extensively
studied (see Refs.~\cite{bigelow_92,metcalf_93,drewsen_95} for example).

\begin{figure}
\includegraphics[width=3in]{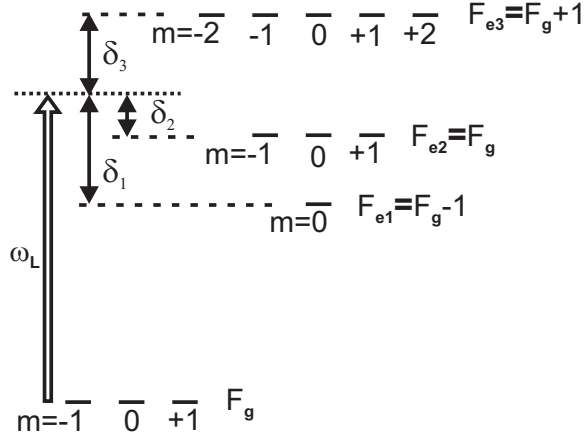}
\caption{\label{fig:multlevel}Energy level diagram of an atom with multiple hyperfine manifolds. If
the energy spacing of the excited-state manifolds are of the order or smaller than the natural
linewidth of the transition, the usual sub-Doppler cooling transition ($F_g \leftrightarrow F_e =
F_g+1$) is not isolated and the other manifolds must be taken into account.}
\end{figure}

This paper focuses primarily on monochromatic laser cooling for atoms with multiple closely spaced
hyperfine excited-state manifolds. Figure~\ref{fig:multlevel} provides a graphical illustration of
this type of atomic configuration. This situation is of importance, for example, in alkaline-earth
atoms with nonzero nuclear magnetic moment. If the excited state manifolds are spaced in energy on
the order or smaller than the excited state linewidth $\gamma$, coherences between these manifolds
become nonnegligible, and can have a significant effect on the optical pumping processes required
for sub-Doppler cooling and on the dynamics of the atom-photon interaction. Sub-Doppler laser
cooling was experimentally identified in fermionic $^{87}$Sr~\cite{me_03}, despite significant
spectral overlap in the excited state. At the time, it was hypothesized that the large ground-state
degeneracy in $^{87}$Sr (due to the large nuclear spin $I=9/2$) was somehow able to overcome the
decrease in cooling due to the spectral overlap. Other systems with spectral overlap in the excited
state are $^{39}$K~\cite{bambini_98}, $^{7}$Li~\cite{grimm_98}, and the fermionic isotopes of
Yb~\cite{fortson_03}. In $^{87}$Rb, the effects of excited-state spectral overlap on the
effectiveness of velocity-selective coherent population trapping have been explored, both
experimentally and theoretically~\cite{grynberg_99}. Our goal in the paper is to provide a detailed
discussion of the theoretical techniques required to model such systems realistically. In a future
publication, we plan to present comprehensive laser-cooling predictions for a variety of atoms.

The structure of this paper is as follows. In Section II, we develop the master equation for a
laser-driven atom with multiple excited-state manifolds, and then specialize this equation to the
low-intensity limit. In Section III, we introduce the MCWF technique and apply it to this
low-intensity master equation. In Section V, we perform full Monte Carlo master-equation
simulations for $^{25}$Mg and $^{87}$Sr atoms in a 3D lin-$\perp$-lin laser configuration as an
example of using this technique determine expected temperatures for these atoms in a laser cooling
experiment. In Section VI, we conclude.

\section{Master Equation in the Low-Intensity Limit}

In this section we develop the master equation describing a multilevel atom interacting with a
coherent laser field and coupled to a vacuum photon field. It is this equation, with quantized
atomic CM, that will provide an accurate description of atom-photon dynamics, and this master
equation will provide the basis for the Monte Carlo simulations that will be discussed later.

The full Hamiltonian for the atom-laser system plus the radiation field is
 \begin{equation}
 \label{eq:full_ham}
 H = H_{A} + H_{R} + V_{A-L} + V_{A-R},
 \end{equation}
where $H_{A} = \sum_{i} \hbar \omega_{i} P_{i} + \frac{P^2}{2m}$ is the bare atomic Hamiltonian,
$H_{R}$ is the vacuum radiation field Hamiltonian, and $V_{A-L}$ and $V_{A-R}$ are the atom-laser
and atom-radiation field coupling terms, respectively. In the atomic Hamiltonian, $P_{i}$ is a
projection operator onto the $i$-th internal excited-state manifold, $\hbar \omega_{i}$ is the
energy of the $i$-th excited-state manifold relative to the ground-state manifold, $P$ is the
atomic CM momentum operator, $m$ is the atomic mass, and the sum runs over all excited-state
manifolds. We have assumed in Eq.~(\ref{eq:full_ham}) that the effects of atom-laser and
atom-radiation-field coupling are independent~\cite{cohen_tannoudji_book_92}.

We can view Eq.~(\ref{eq:full_ham}) in terms of system-reservoir interactions. The \emph{system}
consists of the atom, the laser, and their interaction. The system Hamiltonian is
 \begin{equation}
 H_S = H_A + V_{A-L}.
 \end{equation}
The \emph{reservoir} is the vacuum radiation field, having many more modes than the system. With
the Markov approximation, along with a few other approximations, the master equation is then given
by
 \begin{equation}
 \label{eq:full_master_eqn}
 \sigd = \frac{i}{\hbar}\left[\sigma,H_S\right] + \lsp[\sigma].
 \end{equation}
The operator $\sigma$ is the system reduced density operator element, i.e., the reservoir degrees
of freedom have been traced over, $\sigma=\text{Tr}_R\rho$. The remaining term, $\lsp[\sigma]$,
encompasses the interaction between the atom and the vacuum photon field, and provides for the
phenomenon of spontaneous emission.

The relaxation operator due to spontaneous emission, which we derive in detail in the Appendix, is
given by
 \begin{multline}
 \label{eq:full_lsp}
 \lsp[\sigma]
 =
 \frac{3\gamma}{8\pi}
 \int d^2\Omega \sum_{\bm{\epsilon}\perp\mathbf{k}}
 \sum_{i,j}
 e^{-i \mathbf{k}\cdot\mathbf{R}} ( \bm{\epsilon}^{*} \cdot \mathbf{A}^{(i)} )
 \sigma
 (\bm{\epsilon}\cdot{\mathbf{A}^{(j)}}^{\dagger}) e^{i \mathbf{k}\cdot\mathbf{R}}
 \\
 -
 \frac{3\gamma}{16\pi}
 \int d^2\Omega \sum_{\bm{\epsilon}\perp\mathbf{k}}
 \sum_{i,j}
 \left[
 (\bm{\epsilon}\cdot{\mathbf{A}^{(i)}}^{\dagger}) e^{ i \mathbf{k}\cdot\mathbf{R}}
 e^{-i \mathbf{k}\cdot\mathbf{R}} (\bm{\epsilon}^{*}\cdot\mathbf{A}^{(j)})
 \sigma
 +
 \sigma
 (\bm{\epsilon}\cdot{\mathbf{A}^{(i)}}^{\dagger}) e^{ i \mathbf{k}\cdot\mathbf{R}}
 e^{-i \mathbf{k}\cdot\mathbf{R}} (\bm{\epsilon}^{*}\cdot\mathbf{A}^{(j)})
 \right],
 \end{multline}
where ${\mathbf{A}^{(i)}}^{\dagger}$ and $\mathbf{A}^{(i)}$ are vector raising and lowering
operators, respectively, between the ground state and the $i$th excited state, $\mathbf{R}$ is the
atomic CM position, $\mathbf{k}$ is the direction of the photon emitted in the relaxation process,
and $\gamma$ is the decay rate of the exited states. The integral is performed over solid angle in
the vector $\mathbf{k}$ and the sum over $\bm{\epsilon}\perp\mathbf{k}$ refers to the two
polarization directions perpendicular to $\mathbf{k}$. Note that here and throughout this paper, we
assume that each of the excited-state hyperfine manifolds has the same lifetime $\tau =
\gamma^{-1}$. Expanding these vector operators in a basis of spherical unit vectors,
$\hat{\epsilon}_{\pm 1} = \mp(\hat{x} \pm i\hat{y})/\sqrt{2}$ and $\hat{\epsilon}_{0} = \hat{z}$,
we have
 \begin{equation}
 \mathbf{A}^{(i)} = \sum_{q=0,\pm 1} (-1)^{q} \hat{\epsilon}_{-q} A_q^{(i)}.
 \end{equation}
The spherical components of the vector operators are
 \begin{align}
 \label{eq:adefs}
 A^{(i)}_q
 &=
 \sum_{M_g,M_{e_i}}
 \alpha_{F_g,F_{e_i},M_g,M_{e_i},J_g,J_e,I}
 \ket{J_g I F_g M_g}\bra{J_e I F_{e_i} M_{e_i}}\\
 {A^{(i)}_q}^{\dagger}
 &=
 \sum_{M_g,M_{e_i}}
 \alpha_{F_g,F_{e_i},M_g,M_{e_i},J_g,J_e,I}
 \ket{J_e I F_{e_i} M_{e_i}}\bra{J_g I F_g M_g},
 \end{align}
where
 \begin{multline}
 \alpha_{F_g,F_{e_i},M_g,M_{e_i},J_g,J_e,I}
 =
 (-1)^{F_g + F_{e_i} + M_g + J_e + I}
 \sqrt{(2 F_g + 1)(2 F_{e_i} + 1)(2 J_e + 1)}\\
 \times
 \begin{pmatrix}
  F_g & 1 & F_{e_i}\\
  -M_g & M_g - M_{e_i} & M_{e_i}
 \end{pmatrix}
 \begin{Bmatrix}
  J_g & F_g & I\\
  F_{e_i} & J_e & 1
 \end{Bmatrix}.
 \end{multline}

Eq.~(\ref{eq:full_lsp}) is written in a way that makes explicit that it is in Lindblad
form~\cite{meystre_book_98,zoller_book_04,spohn_80}. As we will see later, it is important for the
relaxation operator to be of this form in order to make use of the MCWF technique. Because the
complex exponentials in the second line cancel each other, the remaining integral over solid angle
can be evaluated, whereby Eq.~(\ref{eq:full_lsp}) can be equivalently written as~\cite{molmer_93}
 \begin{equation}
 \label{eq:lsp_2}
 \lsp[\sigma]
 =
 \frac{3\gamma}{8\pi}
 \int d^2\Omega \sum_{\bm{\epsilon}\perp\mathbf{k}}
 \sum_{i,j}
 e^{-i \mathbf{k}\cdot\mathbf{R}} ( \bm{\epsilon}^{*} \cdot \mathbf{A}^{(i)} )
 \sigma
 (\bm{\epsilon}\cdot{\mathbf{A}^{(j)}}^{\dagger}) e^{i \mathbf{k}\cdot\mathbf{R}}
 -\frac{\gamma}{2} \sum_i \left[P_{e_i}\sigma + \sigma P_{e_i}\right].
 \end{equation}

We will now examine that atom-laser interaction term, which is given in the electric-dipole
approximation by
 \begin{equation}
 V_{A-L}(\mathbf{R},t) = -\mathbf{D} \cdot \mathbf{E}_{L}(\mathbf{R},t),
 \end{equation}
where $\mathbf{E}_{L}(\mathbf{R},t)$ is the electric field of the laser and $\mathbf{D}$ is the
electric dipole operator. As usual, we treat the laser as a classical field, since it is a densely
populated mode of the electric field. We can write the laser electric field in terms of its
positive and negative frequency components, $\mathbf{E}_{L}(\mathbf{R},t) =
\mathbf{E}_{L}^{(+)}(\mathbf{R}) e^{-i \omega t} + \text{c.c.}$, and then expand into spherical
components,
 \begin{equation}
 \mathbf{E}_{L}^{(+)}(\mathbf{R})
 =
 \frac{E_0}{2} \sum_{q=0,\pm 1} (-1)^{q} a_q(\mathbf{R}) \hat{\epsilon}_{-q},
 \end{equation}
where $E_0$ is the electric-field amplitude and $a_q(\mathbf{R})$ are the expansion coefficients.
Making the rotating-wave approximation, so that
 \begin{equation}
 V_{A-L}(\mathbf{R},t)
 =
 - \mathbf{D}^{(+)} \cdot \mathbf{E}_{L}^{(+)}(\mathbf{R}) e^{-i \omega t}
 - \mathbf{D}^{(-)} \cdot \mathbf{E}_{L}^{(-)}(\mathbf{R}) e^{ i \omega t},
 \end{equation}
where $\mathbf{D}^{(+)} = \sum_i P_{e_i} \mathbf{D} P_g$ and $\mathbf{D}^{(-)} = \sum_i P_g
\mathbf{D} P_{e_i}$, we find
 \begin{equation}
 \label{eq:val}
 V_{A-L} = -\frac{\Omega}{2} \sum_i \mathcal{D}_i(\mathbf{R})e^{-i\omega t} + \text{H.c.}.
 \end{equation}
In the previous equation we have defined the atom-laser raising operator,
 \begin{equation}
 \label{eq:drld}
 \mathcal{D}_i^{\dagger}(\mathbf{R})
 =
 \sum_{q=0,\pm 1} a_q(\mathbf{R}) {A_{q}^{(i)}}^{\dagger},
 \end{equation}
and lowering operator,
 \begin{equation}
 \label{eq:drl}
 \mathcal{D}_i(\mathbf{R})
 =
 \sum_{q=0,\pm 1} a_q^{*}(\mathbf{R}) A_{q}^{(i)},
 \end{equation}
and introduced the "invariant" Rabi frequency,
 \begin{equation}
 \Omega = \frac{E_{0}\braket{J_e||D||J_g}}{\sqrt{2 J_e + 1}},
 \end{equation}
where $\braket{J_e||D||J_g}$ is the reduced dipole matrix element between the ground and excited
states. This form of a Rabi frequency, defined in terms of the reduced matrix element between the
$J=J_g$ ground state and the $J=J_e$ excited state, is convenient because, in general, Rabi
frequencies for transitions to different excited-state manifolds will not be the same.

Next, we observe that the second term in Eq.~(\ref{eq:lsp_2}) is comprised of excited-state
projection operators both pre- and post-multiplying the system density operator. Thus, it is clear
that this term can be absorbed into the free-evolution commutator term in
Eq.~(\ref{eq:full_master_eqn}), allowing the master equation to be equivalently described by
Hamiltonian evolution determined by an effective Hamiltonian $H_{\text{eff}}$, plus a term which is
commonly called a \emph{jump term}, and which cannot be written in the form of a commutator with
the system density operator. We thus have,
 \begin{equation}
 \label{eq:me_gen_heff}
 \dot{\sigma}
 =
 -\frac{i}{\hbar}\left(H_\text{eff} \sigma - \sigma H_\text{eff}^{\dagger}\right)
 +
 \frac{3\gamma}{8\pi}
 \int d^2\Omega \sum_{\bm{\epsilon}\perp\mathbf{k}}
 \sum_{i,j}
 e^{-i \mathbf{k}\cdot\mathbf{R}} (\bm{\epsilon}^{*}\cdot\mathbf{A}^{(i)})
 \sigma
 (\bm{\epsilon}\cdot{\mathbf{A}^{(j)}}^{\dagger}) e^{i \mathbf{k}\cdot\mathbf{R}},
 \end{equation}
where the effective Hamiltonian $H_\text{eff}$ is given by
 \begin{equation}
 \label{eq:gen_heff}
 H_\text{eff}
 =
 \frac{P^{2}}{2m} - \sum_i \hbar\left(\delta_i + i\frac{\gamma}{2}\right)P_{e_i} + V_{A-L},
 \end{equation}
where $V_{A-L}$ is as given in Eq.~(\ref{eq:val}). In obtaining Eqs.~(\ref{eq:me_gen_heff})
and~(\ref{eq:gen_heff}), we have made the usual rotating-frame transformation, which removes the
free-evolution atomic Bohr frequencies from the problem. The more relevant frequencies are instead
the laser detunings $\delta_i = \omega - \omega _{i}$ from the $i$th excited-state hyperfine
manifold. The master equation given in Eq.~(\ref{eq:me_gen_heff}) is fully general, but has been
written in a form that will facilitate setting up a stochastic wave function simulation using the
MCWF technique described later.

We would like to now specialize the master equation just discussed to the limit of low laser
intensity. Specifically, this limit is valid when the saturation parameter for the atom in the
$i$th excited-state hyperfine manifold,
 \begin{equation}
 s_i = \frac{\Omega^2 / 2}{\delta_i^2 + (\gamma/2)^2},
 \end{equation}
is small, which occurs when the laser intensity is small or the laser detuning from the atomic
transition is large. In this limit, the excited states are said to adiabatically follow the ground
states. The excited states can then be eliminated from the equations of motion, resulting in a
master equation in terms of only the ground-state sub-density-matrix,
 \begin{equation}
 \sigma_{gg} = P_{g} \sigma P_{g}.
 \end{equation}
In this limit, the master equation becomes (see section 8.3.3 of
Ref.~\cite{cohen_tannoudji_proceedings_92})
 \begin{equation}
 \dot{\sigma}_{gg}
 =
 -\frac{i}{\hbar}\left(h_\text{eff} \sigma_{gg} - \sigma_{gg} h_\text{eff}^{\dagger}\right)
 +
 \int d^2\Omega \sum_{\bm{\epsilon}\perp\mathbf{k}}
 \sum_{i,j}
 ( \bm{\epsilon}^{*} \cdot \mathbf{B}^{(i)}(\mathbf{R},\mathbf{k}) )
 \sigma_{gg}
 ( \bm{\epsilon} \cdot {\mathbf{B}^{(i)}}^{\dagger}(\mathbf{R},\mathbf{k}) ).
 \end{equation}
The new effective Hamiltonian is given by
 \begin{equation}
 \label{eq:liheff}
 h_\text{eff}
 =
 \frac{P^{2}}{2m}
 + \sum_i \frac{s_i}{2} \hbar\left(\delta_i - i\frac{\gamma}{2}\right)
 \mathcal{D}^{(i)}(\mathbf{R}){\mathcal{D}^{(i)}}^{\dagger}(\mathbf{R}).
 \end{equation}
The new decay raising and lowering operators are given by
 \begin{equation}
 \label{eq:bqd}
 {B_{q}^{(i)}}^{\dagger}(\mathbf{R},\mathbf{k})
 =
 \sqrt{\frac{3 s_i \gamma}{8\pi}}
 {A_{q}^{(i)}}^{\dagger} e^{i \mathbf{k}\cdot\mathbf{R}} \mathcal{D}^{(i)}(\mathbf{R}),
 \end{equation}
and
 \begin{equation}
 \label{eq:bq}
 B_{q}^{(i)}(\mathbf{R},\mathbf{k})
 =
 \sqrt{\frac{3 s_i \gamma}{8\pi}}
 A_{q}^{(i)} e^{-i \mathbf{k}\cdot\mathbf{R}} {\mathcal{D}^{(i)}}^{\dagger}(\mathbf{R}).
 \end{equation}
Note that this new lowering (raising) operator contains two components: a raising (lowering)
operator ${\mathcal{D}^{(i)}}^{\dagger}(\mathbf{R})$ (${\mathcal{D}^{(i)}}(\mathbf{R})$) between
the ground state and the $i$th excited-state manifold due to the atom-laser interaction, and a
lowering (raising) operator ${A_{q}^{(i)}}^{\dagger} e^{i \mathbf{k}\cdot\mathbf{R}}$
(${A_{q}^{(i)}} e^{-i \mathbf{k}\cdot\mathbf{R}}$) of type $q$ corresponding to coupling with the
reservoir photon field via a photon with polarization $q$. Thus, the jump operator in the
low-intensity equations describes a transition cycle of the atom involving coupling to both the
laser and the reservoir photon field. Note also that this new operator and the
effective-Hamiltonian term in the equation of motion are both proportional to the saturation
parameter $s_i$, the perturbation parameter.

\section{The Monte Carlo Wave-Function Technique}

The MCWF~\cite{molmer_92,molmer_93,molmer_95,zoller_92,percival_92,percival_92b,carmichael_book_93}
technique is a means of interpreting a system-reservoir master equation --- which describes the
evolution of a density operator for a system interacting with a large external reservoir --- as the
evolution of an ensemble of individual wave functions, each undergoing random \emph{quantum jumps}.
The free evolution of the stochastic wave functions is determined by the effective Hamiltonian that
we found in the previous section. The nature of the quantum jumps is determined by the leftover
term in the master equation, which cannot be absorbed into the free-evolution commutator. The
components of this leftover term are often called quantum-jump operators.

In the following, we will deal primarily with the master equation in the low-intensity limit, as
developed in the previous section, although the methods could just as easily be applied to the
arbitrary-intensity master equation. The low-intensity limit, however, provides a reduction in the
number of internal atomic states required in the calculation, and this will be beneficial for
performing calculations later. Furthermore, since the lowest temperatures are achieved for low
laser intensities, such a specialization does not hinder our ability to calculate lower bounds of
temperature.

Having already expressed the master equation in a form involving an effective Hamiltonian and a
jump term in the previous section, the application of the MCWF technique is rather straightforward
along the lines developed in the literature (see, in particular, Ref.~\cite{molmer_93}). For a
single stochastic wave function, the procedure is as follows. First, set the wave function to an
initial value. Then, numerically propagate the wave function for a time step $\delta t$ according
to the effective Hamiltonian $H_\text{eff}$ only, from an initial value $\ket{\psi(t)}$ to a final
value $\ket{\psi^{(1)}(t + \delta t)}$,
 \begin{equation}
 \label{eq:heffevolve}
 \ket{\psi^{(1)}(t + \delta t)}
 =
 \left( 1 - \frac{i H_\text{eff} \delta t}{\hbar} \right) \ket{\psi(t)}.
 \end{equation}
Restrictions on the size of $\delta t$ are given such that the first-order truncation of the
time-evolution operator in Eq.~(\ref{eq:heffevolve}) is approximately valid. We note that
$H_\text{eff}$ is non-Hermitian by construction, as a result of absorbing parts of the relaxation
operator into the original (Hermitian) bare system Hamiltonian. Because of this, propagation with
$H_\text{eff}$ will not conserve the norm of the wave function when propagated to
$\ket{\psi^{(1)}(t + \delta t)}$. The time step $\delta t$ of the propagation must be chosen so
that $\delta p \ll 1$ in the inner product,
 \begin{equation}
 \Braket{\psi^{(1)}(t + \delta t) | \psi^{(1)}(t + \delta t)} = 1 - \delta p.
 \end{equation}
The quantity $\delta p$ is the loss of norm resulting from propagating with $H_\text{eff}$ for a
time step $\delta t$, and is found to be
 \begin{equation}
 \begin{split}
 \delta p
 &=
 \delta t
 \braket{\psi(t)
 |
 \sum_i {\mathbf{B}^{(i)}}^{\dagger}(\mathbf{R},\mathbf{k}) \cdot \mathbf{B}^{(i)}(\mathbf{R},\mathbf{k})
 |
 \psi(t)}\\
 &=
 \delta t
 \braket{\psi(t)
 |
 \sum_i \sum_{q=0,\pm 1} {B_{q}^{(i)}}^{\dagger}(\mathbf{R},\mathbf{k}) B_{q}^{(i)}(\mathbf{R},\mathbf{k})
 |
 \psi(t)}\\
 &=
 \sum_i \sum_{q=0,\pm 1} \delta p_{i,q}.
 \end{split}
 \end{equation}
The total loss of norm has been decomposed into individual elements each corresponding to a
particular type of interaction with the reservoir (i.e., the $q$-value of the interaction, or the
excited state $i$ involved). These individual contributions are given by
 \begin{equation}
 \delta p_{i,q}
 =
 \delta t
 \braket{\psi(t)
 |
 {B_{q}^{(i)}}^{\dagger}(\mathbf{R},\mathbf{k}) B_{q}^{(i)}(\mathbf{R},\mathbf{k})
 |
 \psi(t)}.
 \end{equation}
We see that the loss of norm due to a given type of interaction with the reservoir is determined by
the quantum-mechanical expectation value of the product of jump operators of this type of
interaction. The loss of norm $\delta p$ can also be interpreted as the probability for a quantum
jump to occur.

After the wave function has been propagated as described above, and the values of $\delta p_{i,q}$
calculated, it must then be determined whether or not a quantum jump occurred. This is achieved by
generating a pseudo-random number on a computer and comparing it to the value of the total jump
probability $\delta p$. If the random number is less than $\delta p$, a quantum jump occurred, and
if it is greater, no quantum jump occurred. If a quantum jump does occur, the type of quantum jump
must also be calculated by comparing the random number with the individual sub-probabilities
$\delta p_{i,q}$ in the same manner.

If a quantum jump of type $q,i$ occurs, we must apply the quantum jump lowering operator
$B_{q}^{(i)}(\mathbf{R},\mathbf{k})$ to the wave function from the beginning of the time step,
 \begin{equation}
 \ket{\psi(t + \delta t)}
 =
 \sqrt{\frac{\delta t}{\delta p_{i,q}}}
 B_{q}^{(i)}(\mathbf{R},\mathbf{k})
 \ket{\psi(t)}.
 \end{equation}
The square-root factor in front of the lowering operator is necessary for renormalization. If no
quantum jump occurs, then we simply renormalize the wave function.

The resulting wave function is then used as the starting point for propagation over the next time
step, and the procedure is repeated.

A good approximation of the true system density matrix is achieved by combining the trajectories of
a number of independently propagated stochastic wave functions, each trajectory having a unique
sequence of pseudo-random numbers. (A thorough discussion of the statistical issues involved with
the MCWF technique can be found in Ref.~\cite{molmer_93}.) Once a suitable ensemble of stochastic
wave function trajectories has been obtained, an estimate of the true expectation value of an
operator is found by taking the ensemble average of the expectation value of that operator with
respect to the stochastic wave functions. For example, an estimate of the average kinetic energy at
a time $t$ for a system for which $N$ independent stochastic wave functions have been calculated is
given by
\begin{equation}
 \left<E\right>(t) = \frac{1}{N}
 \sum_{i=1}^{N} \Braket{\psi_i (t) | \frac{P^2}{2m} | \psi_i (t)},
\end{equation}
where $\psi_i (t)$ is the $i$th stochastic wave function, given at time $t$.

\begin{figure}
\includegraphics[width=3.5in]{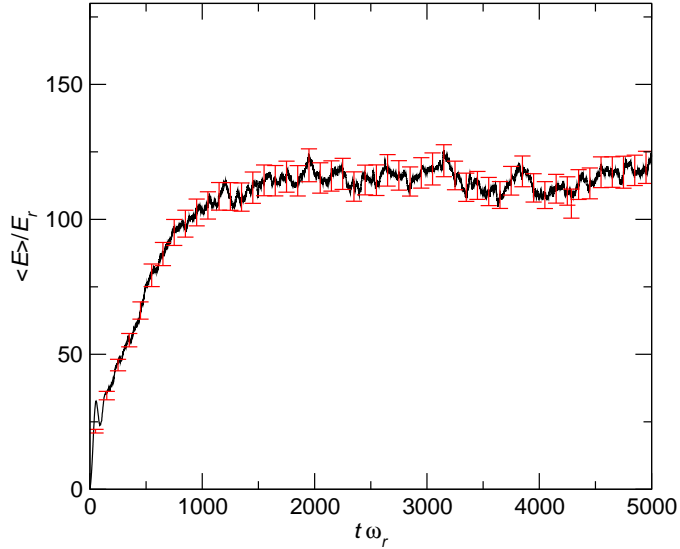}
\caption{\label{fig:wfs}(Color online) An example of a characteristic MCWF stochastic trajectory.
Shown is the average of the atomic CM energy over 500 independent stochastic wave functions, as a
function of time, for a two-level atom in a 1D standing-wave laser field. The energy is given in
units of the recoil energy $E_r = \hbar^2 k^2 / 2m$, and time is given in units of the inverse
recoil frequency $\omega_r^{-1} = \hbar/E_r$. All wave functions are initialized in the ground
state of the atom and localized in momentum space with zero momentum. The steady state, wherein the
system fluctuates around an average value, is seen to be achieved after a transient relaxation
period. Error bars indicate the variance in the data at each given time for the ensemble of 500
stochastic wave functions. An estimate of the steady-state atomic CM energy is obtained by
performing a time-average over all wave functions for all times after the relaxation regime.  The
error bar of such an average will be smaller than the error bars in the figure, which apply only to
the data for a given time.}
\end{figure}

Figure~\ref{fig:wfs} demonstrates a simple example of the application of the MCWF technique,
wherein the average kinetic energy is calculated for a two-level atom interacting with a
one-dimensional standing-wave field. For this calculation, we have used a Rabi frequency of $\Omega
= \gamma/2$ and a detuning of $\delta = -\gamma/2$, where $\gamma$ is the decay rate of the upper
to the lower atomic state, and we have set $\gamma = 400 E_r$, where $E_r = \hbar^2 k^2 / 2m$ is
the recoil energy. The atomic kinetic energy, averaged over 500 stochastic wave functions each
initialized to zero momentum, is plotted as a function of time, with error bars indicating the
error in the ensemble average for a given time. The separation of the transient relaxation period
from the steady-state is clear, the steady state regime being characterized by fluctuations in the
average energy about a mean. This noise is due to the finite number of stochastic wave functions
being used, and if a greater number of wave functions were used, the amplitude of the fluctuations
would be decreased. In the limit of an infinite number of wave functions, the true density-matrix
solution of the master equation would be obtained. An estimate of the steady-state kinetic energy
is found by time-averaging the calculated data over the entire steady-state regime. Since this is a
larger ensemble than the set of wave functions for a single time, the error of such an average will
be smaller than the error bars shown in the figure.

\section{Calculations for $^{25}$Mg and $^{87}$Sr}

The purpose of this section is to illustrate the application of the theory developed up to this
point to a complicated system. We wish to quantitatively study the dynamics of particular atoms
interacting with 3D polarization-gradient laser fields. The balance of the frictional cooling
forces along with the diffusion experienced by the atom due to spontaneous emission and its
interaction with the laser leads to a steady-state momentum distribution that determines the
temperature of a gas of such atoms. In particular, we will study here the cooling of the fermionic
isotopes of two alkaline-earth atoms, $^{25}$Mg (nuclear spin $I=5/2$, $^{1}S_{0}$-$^{1}P_{1}$
width $\gamma/2\pi =$ 81$\,$MHz, hyperfine splittings $\Delta \omega_{13}/2\pi =$ 46$\,$MHz and
$\Delta \omega_{23}/2\pi =$ 27$\,$MHz, where we have assumed a hyperfine quadrupole parameter
$B=0$~\cite{kluge_74}) and $^{87}$Sr ($I=9/2$, $^{1}S_{0}$-$^{1}P_{1}$ width $\gamma/2\pi =$
32$\,$MHz, hyperfine splittings $\Delta \omega_{13}/2\pi =$ 43$\,$MHz and $\Delta \omega_{23}/2\pi
=$ -17$\,$MHz). These atoms, having nonzero nuclear magnetic moment, have degenerate (assuming zero
magnetic field) Zeeman sublevels. These sublevels allow for the mechanism of sub-Doppler cooling in
an appropriate laser configuration. Both $^{25}$Mg and $^{87}$Sr exhibit significant excited-state
spectral overlap, with $\Delta \omega_{13}/\gamma =$ 0.57, $\Delta \omega_{23}/\gamma =$ 0.33, and
$\Delta \omega_{13}/\gamma =$ 1.3, $\Delta \omega_{23}/\gamma =$ -0.53, respectively. We consider
the 3D lin-$\perp$-lin laser configuration, consisting of a pair of opposing beams along each
cartesian axis, in which each beam is linearly polarized orthogonal to its opposing beam.
Furthermore, for this calculation, we set to zero the relative phases of the three sets of laser
pairs.

Having a nuclear spin of $I=5/2$, the $^{1}S_0$ state of $^{25}$Mg results in a hyperfine ground
state with 6 sublevels. Use of the low-intensity master equation given in
Eq.~(\ref{eq:me_gen_heff}) allows us to consider only these 6 internal states of the atom, since
the excited states have been adiabatically eliminated in this regime. However, as noted in
Ref.~\cite{molmer_95}, a momentum grid extending to $20 \hbar k$ in each direction with a spacing
of $\hbar k$ would yield a density matrix with $(6 \times 41^3)^2 \approx 2 \times 10^{11}$
elements. A direct solution of this master equation is not numerically feasible, even without
considering the further increases in matrix size necessary to describe the master equation
relaxation operator in Liouville space~\cite{blum_book_81}. On the other hand, the MCWF method only
requires numerical propagation of individual wave functions, which would be represented by vectors
with $6 \times 41^3 \approx 4 \times 10^5$ elements. If the number of independent stochastic wave
functions required to achieve satisfactory convergence for the calculation of a particular property
of the system is not unreasonably large, the MCWF method provides a distinct advantage over a
direct master-equation solution.

We follow the procedure outlined in Section III, working in the low-intensity limit in order to
reduce the number of internal atomic states in the calculation, which increases the efficiency of
calculation. Since laser cooling is most effective at low laser intensities, this turns out to be a
useful regime in which to work, with the additional benefit that lower temperatures require a
smaller number of atomic CM momentum states in the calculation. We must determine the effective
Hamiltonian as given in Eq.~(\ref{eq:liheff}) and the jump operators as given in
Eqs.~(\ref{eq:bqd}) and~(\ref{eq:bq}) for each atom, and for the particular laser field being
considered.

We consider here the lin-$\perp$-lin laser configuration in 3D, with the relative phases of the
beams set to zero. The positive-frequency component of the electric field is
 \begin{equation}
 \begin{split}
 \mathbf{E}_{L}(\mathbf{R},t)
 &=
 \frac{E_0}{2}
 \left[
 \hat{y} e^{i k X} + \hat{z} e^{-i k X}
 +
 \hat{z} e^{i k Y} + \hat{x} e^{-i k Y}
 +
 \hat{x} e^{i k Z} + \hat{y} e^{-i k Z}
 \right]\\
 &=
 \frac{E_0}{2}
 \sum_{q=0,\pm 1} (-1)^q a_q (\mathbf{R}) \hat{\epsilon}_{-q},
 \end{split}
 \end{equation}
with spherical coefficients
 \begin{gather}
 a_{+1}(\mathbf{R}) = -\frac{1}{\sqrt{2}}
 \left( e^{-i k Y} + e^{i k Z} + i e^{i k X} + i e^{-i k Z} \right),\\
 a_{-1}(\mathbf{R}) = +\frac{1}{\sqrt{2}}
 \left( e^{-i k Y} + e^{i k Z} - i e^{i k X} - i e^{-i k Z} \right),\\
 a_{0}(\mathbf{R}) =
 e^{-i k X} + e^{i k Y}.
 \end{gather}
With these coefficients, along with parameters appropriate to the particular atom under
consideration, the atom-laser raising and lowering operators given in Eqs.~(\ref{eq:drld})
and~(\ref{eq:drl}) can be constructed. With knowledge of the effective Hamiltonian and the raising
and lowering operators, we can then proceed with the MCWF procedure as outlined.

\begin{figure}
\includegraphics[width=3.5in]{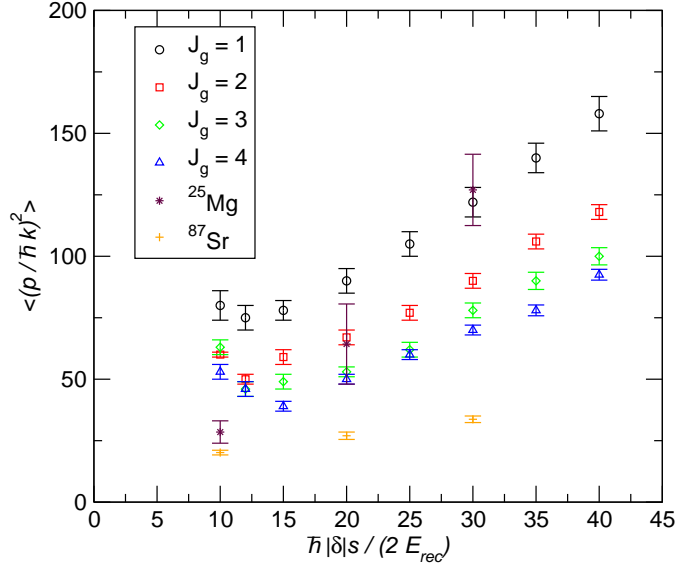}
\caption{\label{fig:temps} (Color online) Results for calculated ensemble-average energies (rms
momentum squared) for $^{25}$Mg and $^{87}$Sr, as a function of the light-shift parameter $\hbar
|\delta_3| s_3 / (2 E_\text{rec})$.  For comparison, also shown are the calculated energies for
atoms with isolated transitions, $J_e = J_g + 1$, with $J_g =$1, 2, 3, and 4, with detuning $\delta
= -5\gamma$.  See text for discussion.}
\end{figure}

Our example entails propagating 20 stochastic wave functions each for three different values of the
light-shift parameter, $\hbar |\delta_3| s_3 / (2 E_\text{rec}) =$10, 20, and 30, for both
$^{25}$Mg ($I=5/2$) and $^{87}$Sr ($I=9/2$). We consider only $\delta_3 = -5 \gamma$. As in
Figure~\ref{fig:wfs}, we calculate the stochastic trajectories of the ensemble average (i.e.,
averaged over the 20 wave functions) kinetic energy for each atom as a function of time. We
continue this propagation until the transient regime has been passed for some time, and use the
time average over the steady-state ensemble-average kinetic energy to provide an estimate of the
total average kinetic energy and the final error. The results are shown in Fig.~\ref{fig:temps},
along with the energies for atoms with an isolated cooling transition for comparison, $J_e = J_g +
1$ with $J_g =$1, 2, 3, and 4, with detuning $\delta = -5\gamma$, as first calculated by Castin and
M{\o}lmer in Ref.~\cite{molmer_95}. From this cursory analysis, we can see that $^{25}$Mg should
exhibit a sharp rise in temperature with increasing laser intensity, while $^{87}$Sr will cool to
sub-Doppler temperatures even for higher intensities, as has been noted
experimentally~\cite{me_03}.

Detailed calculations of this sort, for realistic atoms, are quite computationally expensive. For
example, a single data point for the Mg and Sr calculations presented here required on the order of
200 hours wall time for a 20 processor parallel code, running on a cluster of 2.4 GHz Intel Zeon
processors. There remains work to be done improving the numerical efficiency of our initial codes.
Our goal in this paper has been to present our method and some illustrative results; in a future
publication we plan to expand upon these initial results using improved, faster codes and present
comprehensive predictions of laser cooling temperatures for a variety of atoms.

\section{Conclusions}

In conclusion, we have provided a detailed description of the fully quantum-mechanical master
equation that describes an atom with multiple internal internal structure interacting with a 3D
polarization-gradient laser field. We have shown how the spontaneous-emission relaxation operator
is generalized for atoms of this type. The MCWF technique has been applied to these equations of
motion, providing a more efficient means of performing calculations for these systems compared to a
full solution of the master equation. A few example calculations have been presented to illustrate
the application of this theory to atomic systems interacting with laser configurations commonly
used in experiments. After making improvements in the efficiency of our codes, we intend to expand
upon this work in a future publication and provide a comprehensive survey of laser cooling
calculations for atoms with multilevel internal structure.

\begin{acknowledgments}
We thank R. Santra and J. Ye for helpful discussions. We acknowledge support from the NSF, and use
of computing resources from the Keck Foundation.
\end{acknowledgments}

\appendix*

\section{Relaxation Operator for an Atom with Multiple Excited-State Hyperfine Manifolds}

In this appendix, we outline the major steps in deriving the spontaneous-emission relaxation
operator for an atom with multiple hyperfine excited-state manifolds. Detailed derivations of this
sort, but including only a single excited state manifold, exist elsewhere in the literature (see,
for example, Ref.~\cite{cohen_tannoudji_proceedings_77}). Our intent here is to highlight the steps
important in generalizing the previous work to include coherences between other exited states. We
will work within the framework of the theory of system-reservoir interactions and follow the
notation of Ref.~\cite{cohen_tannoudji_proceedings_77}.

The total Hamiltonian for an atom coupled to a vacuum radiation field is then given by $H = H_{A} +
H_{R} + V_{A-R},$ where $H_A$ and $H_R$ are the atom and reservoir bare Hamiltonian, respectively,
and $V_{A-R}$ is the atom-reservoir coupling and is given in the the electric-dipole approximation
as
 \begin{equation}
 V_{A-R}
 = -\mathbf{D} \cdot \mathbf{E}(\mathbf{R})
 = - \sum_{q=0,\pm 1} (-1)^{q} D_{q} E_{-q}(\mathbf{R}).
 \end{equation}
Here, $\mathbf{D}$ is the electric dipole operator for the atom, and $\mathbf{E}$ is the
electric-field operator for the photon field, and we have expanded the interaction into its
spherical components. To simplify the formalism, we will begin by ignoring the atomic CM and
setting the position coordinate to be the origin, $\mathbf{R} = 0$. At the end we will then
generalize the equations to include the CM degree of freedom.

In general, the total density operator $\rho$ evolves according to the Liouville equation,
$\rhod(t) = \frac{i}{\hbar}\left[\rho(t),H\right]$. Making the usual assumptions involved in
deriving the master
equation~\cite{cohen_tannoudji_book_92,meystre_book_98,cohen_tannoudji_proceedings_77}, we arrive
at an equation of motion for the reduced density operator of the system $\sigma = \text{Tr}_R
\rho$,
 \begin{multline}
 \label{eq:full_eom}
 \sigd(t)
 =
 \frac{i}{\hbar} \left[\sigma(t),H_A\right]
 - \frac{1}{\hbar^2} \int_{0}^{\infty} d\tau \sum_{q} (-1)^{q}\\
 \times
 \left\{
 g_q(\tau)
 \left[
 D_{q} \; e^{-i H_A \tau / \hbar} \; D_{-q} \; e^{ i H_A \tau / \hbar} \; \sigma(t)
 -
 e^{-i H_A \tau / \hbar} \; D_{q} \; e^{ i H_A \tau / \hbar} \; \sigma(t) \; D_{-q}
 \right]
 + \text{H.c.}
 \right\}.
 \end{multline}
In the previous equation, $g_q(\tau)$ is the two-time correlation function of the reservoir and is
defined as $g_q(\tau) = \text{Tr}_{R} [ \sigma_{R} \tilde{E}_q(\tau) \tilde{E}_q(0) ]$, where the
variables with tildes are operators in the interaction representation, $\tilde{E}_q(t) = e^{i H_R t
/ \hbar} E_q e^{-i H_R t / \hbar}$. We assume that the reservoir is initially a vacuum, so that
$\sigma_R = \ket{0}\bra{0}$.  From this we can see $g_q(\tau) = \sum_{\nu} \left| \Braket{\nu | E_q
| 0} \right|^2 e^{-i \omega_{\nu} t}$, where the kets and bras refer to reservoir states. Note that
$g(\tau)^{*} = g(-\tau)$. The correlation time of the reservoir $\tau_C$ is defined such that
$g(\tau) \rightarrow 0$ for $\tau \gg \tau_C$.

In addition to the above approximations, we will also make the \emph{secular approximation}, which
requires that the equation of motion for each density-matrix element $\sigd_{ij}$ have only terms
involving density-matrix elements $\sigma_{kl}$ on the right-hand side such that $|\omega_{ij} -
\omega_{kl}| \ll \gamma$, where $\omega_{ij} \equiv \omega_i - \omega_j$ and where $\gamma$ is the
order of magnitude of the system-reservoir coupling. In the following, we will consider a system
with a ground state coupled to multiple excited states that are separated in energy of the order or
smaller than $\gamma$. Thus, the ground-excited energy splitting $|\omega_{ge}| \gg \gamma$ will be
a non-secular frequency, while $\omega_{e_i e_j} \sim \gamma$ will be a secular frequency.

\begin{figure}
\includegraphics[width=2.5in]{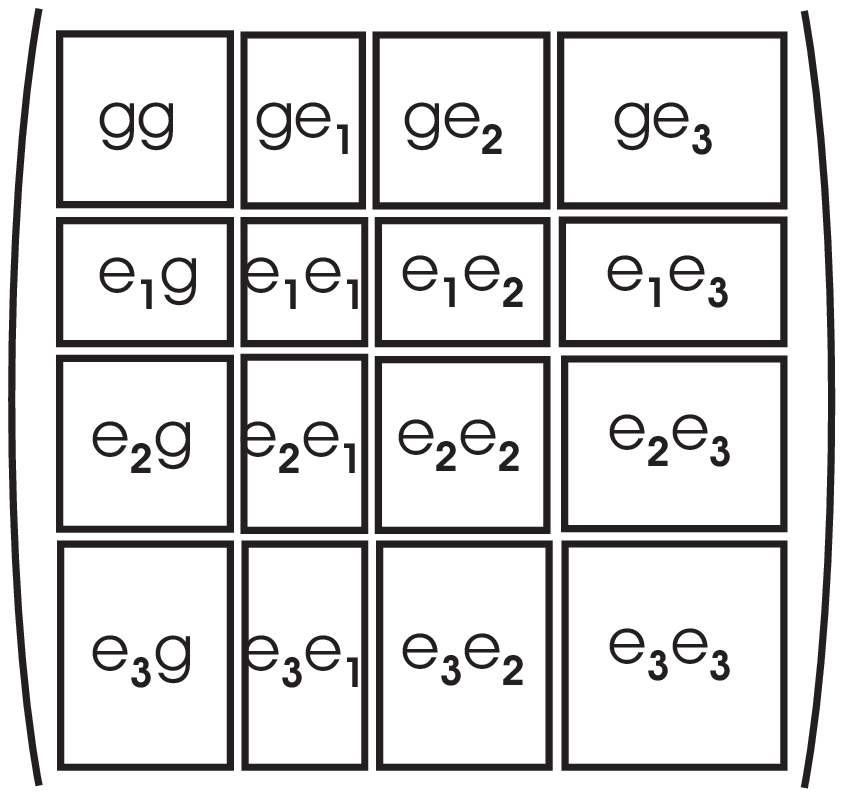}
\caption{\label{fig:dmat}The partitioning of the density operator for an atom with multiple coupled
excited-state manifolds, each potentially having multiple substates.}
\end{figure}

The particular atomic system that we are considering consists of an ground state with electronic
angular momentum $J = J_g = 0$ and an excited state with $J = J_e = 1$. The electronic angular
momentum is coupled to the nuclear spin quantum number $I$, resulting in a ground state with total
angular momentum $F_g = I$, and three excited states with $\{F_{e_i}\} = \{I - 1,I,I + 1\}$. These
assumptions are made for concreteness, but we note that this derivation can be easily extended to
arbitrary angular momentum schemes. It is useful to decompose the system density operator as
illustrated in Fig~\ref{fig:dmat},
 \begin{equation}
 \label{eq:dmat_decomp}
 \sigd(t) = \sigd_{gg}(t) + \sum_{i,j} \sigd_{e_i e_j}(t)
 + \sum_{i} \left[\sigd_{e_i g}(t) + \sigd_{g e_i}(t)\right],
 \end{equation}
where $\sigma_{ij}(t) = P_i \sigma(t) P_j$; $P_i$ is a projection operator onto the $i$-th
hyperfine manifold, $P_i = \sum_{M_i} \ket{J I F M_i}\bra{J I F M_i}$; and $M_i$ is the substate
label for the $i$-th manifold. Two relations that will be useful in the following are
 \begin{gather}
 \sum_{M_g,M_{e_i}}
 \braket{M_g | D_{q} | M_{e_i}} \ket{M_g}\bra{M_{e_i}}
 =
 A^{(i)}_q \frac{\braket{J_g || D || J_e}}{\sqrt{2 J_e + 1}},\\
 \sum_{M_g,M_{e_i}}
 (-1)^{q} \braket{M_{e_i} | D_{-q} | M_g} \ket{M_{e_i}}\bra{M_g}
 =
 {A^{(i)}_q}^{\dagger} \frac{\braket{J_g || D || J_e}^{\dagger}}{\sqrt{2 J_e + 1}},
 \end{gather}
where ${A^{(i)}_q}^{\dagger}$ and $A^{(i)}_q$ are the atomic raising and lowering operators defined
in Eq.~(\ref{eq:adefs}), and where we have made use of symmetry properties of the three-$J$ and
six-$J$ symbols~\cite{sobelman_book_92}.

We focus on the equation for the ground-state sub-density-operator $\sigma_{gg}(t)$ in
Eq.~(\ref{eq:dmat_decomp}). Beginning by taking matrix elements of Eq.~(\ref{eq:full_eom}) between
ground-state sublevel kets, we proceed as usual by eliminating terms that violate energy
conservation (e.g., photon emission coupled to atomic excitation), and absorbing
interaction-induced energy shifts into the energies of the internal atomic levels. The resulting
equation of motion is
 \begin{equation}
 \begin{split}
 \label{eq:sigdgg}
 \sigd_{gg}(t)
 &=
 \frac{{\omega_0}^3}{3 \epsilon_0 (2 \pi)^3 c^3 \hbar}
 \frac{\left|\braket{J_g || D || J_e}\right|^2}{2 J_e + 1}
 \sum_{q} \sum_{i,j}
 A^{(i)}_q \sigma(t) {A^{(j)}_q}^{\dagger}\\
 &\cong
 \gamma
 \sum_{q} \sum_{i,j}
 A^{(i)}_q \sigma(t) {A^{(j)}_q}^{\dagger}.
 \end{split}
 \end{equation}
We have assumed that the energy splittings between ground state and the various excited states are
all approximately equal, and accordingly have defined $\omega_0 = \omega_{e_i} - \omega_{g}$ for $i
=$1,2,3. Equivalently, we have assumed that the decay rate for all of the excited-state manifolds
is approximately equal, and defined $\gamma = \gamma_{J_{e_i} \rightarrow J_g}$ for $i =$,1,2,3.
Note that the double sum over excited-state manifolds in Eq.~(\ref{eq:sigdgg}) will clearly result
in inter-manifold coherence effects in the equations of motion.

Regarding the energy shifts of the internal atomic states that arise due to interaction with the
reservoir states, it is important to mention a subtle feature not found in the simpler case of
degenerate isolated excited states. Such energy shifts occur in the form of divergent
principal-part integrals of virtual transition
amplitudes~\cite{cohen_tannoudji_book_92,cohen_tannoudji_proceedings_77}. For degenerate isolated
manifolds, these diverging terms can be shown to cancel each other in the equations of motion.
However, for the case of multiple, nondegenerate manifolds, these terms no longer cancel exactly,
and pathological divergences related to reservoir-dressed internal atomic energy splittings remain.
A thorough exploration of these terms is outside the scope of this paper, and for the present
purposes, we ignore such diverging terms and absorb interaction-induced energy splittings into the
defined energy levels of the atoms.

Working in the same manner as for the ground-ground sub-density-operator, we can find the equations
of motion for the excited-state sub-density-operators,
 \begin{equation}
 \label{eq:sigdee}
 \sigd_{e_i e_j}(t)
 =
 -i \omega_{e_i e_j} P_{e_i}\sigma(t)P_{e_j}
 - \frac{\gamma}{2} \sum_{q} \sum_{k,l}
 P_{e_i} \left(
 {A^{(k)}_q}^{\dagger} A^{(l)}_q \sigma(t)
 +
 \sigma(t) {A^{(k)}_q}^{\dagger} A^{(l)}_q
 \right) P_{e_j},
 \end{equation}
where we have added a trivial summation index that will be useful later when combining the various
sub-density-matrix decay terms. Similarly, the equations of motion for the optical-coherence
sub-density-operators are
 \begin{equation}
 \label{eq:sigdeg2}
 \sigd_{e_i g}(t)
 =
 -i \omega_{e_i g} P_{e_i}\sigma(t)P_{g}
 - \frac{\gamma}{2} \sum_{q} \sum_{k,l}
 P_{e_i}
 {A^{(k)}_q}^{\dagger} A^{(l)}_q \sigma(t)
 P_{g},
 \end{equation}
and $\sigd_{g e_i}(t) = \sigd_{e_i g}^{\dagger}(t)$.

Using Eq.~(\ref{eq:dmat_decomp}), we can construct the equation of motion due to spontaneous
emission for the full density operator,
 \begin{equation}
 \sigd(t)
 =
 \frac{i}{\hbar} \left[\sigma(t),H_A\right]
 +
 \gamma
 \sum_{q} \sum_{i,j}
 A^{(i)}_q \sigma(t) {A^{(j)}_q}^{\dagger}
 -
 \frac{\gamma}{2}
 \sum_{q} \sum_{i,j}
 \left(
 {A^{(i)}_q}^{\dagger} A^{(j)}_q \sigma(t)
 +
 \sigma(t) {A^{(i)}_q}^{\dagger} A^{(j)}_q
 \right).
 \end{equation}
Defining the spontaneous emission relaxation operator,
 \begin{equation}
 \lsp[\sigma]
 =
 \gamma
 \sum_{q} \sum_{i,j}
 A^{(i)}_q \sigma(t) {A^{(j)}_q}^{\dagger}
 -
 \frac{\gamma}{2}
 \sum_{q} \sum_{i,j}
 \left(
 {A^{(i)}_q}^{\dagger} A^{(j)}_q \sigma(t)
 +
 \sigma(t) {A^{(i)}_q}^{\dagger} A^{(j)}_q
 \right),
 \end{equation}
we can write the equation of motion as
 \begin{equation}
 \sigd(t)
 =
 \frac{i}{\hbar} \left[\sigma(t),H_A\right]
 +
 \lsp[\sigma].
 \end{equation}
Including the atomic CM dependence that we have been ignoring since the beginning amounts to adding
an integral over momentum states in 3D that should have been included when we inserted atomic
projection operators. With this addition, the full relaxation operator takes the form shown in
Eq.~(\ref{eq:full_lsp}).

\end{document}